\documentclass[aps,prl,twocolumn,groupedaddress]{revtex4-1}
\usepackage{mathptmx}

\newcommand{\leaveout}[1]{}

\usepackage{graphicx}
\usepackage{latexsym}
\usepackage{epsfig}
\usepackage{amssymb}
\usepackage{amsmath}
\usepackage{psfrag}
\usepackage{verbatim}

\newcommand{\ignore}[1]{}

\newtheorem{theorem}{Theorem}
\newtheorem{thm}{Theorem}
\newtheorem{lemma}[theorem]{Lemma}

\newtheorem{claim}[theorem]{Claim}

\newtheorem{definition}[theorem]{Definition}
\newtheorem{prop}[theorem]{Proposition}

\newcommand{\cB}{{\cal B}}

\newcommand{\cD}{{\cal D}}
\newcommand{\cE}{{\cal E}}
\newcommand{\cF}{{\cal F}}

\newcommand{\cI}{{\cal I}}

\newcommand{\cN}{{\cal N}}

\newcommand{\cT}{{\cal T}}

\newcommand{\pr}{{\rm Pr}}

\newcommand{\qed}{\hfill $\Box$}

\newcommand{\EX}{\hbox{\bf E}}

\newcommand{\flip}[2]{#1^{(#2)}}
\newcommand{\Inf}{{\rm{Inf}}}

\newcommand{\I}{{I}}
\newcommand{\Infprod}[2]{{\rm{Infprod}}_{#1,#2}}
\newcommand{\nbd}[2]{N^-_{#1,#2}}
\newcommand{\nbdt}[2]{N^-_{\leq #1,#2}}
\newcommand{\cei}[1]{\lceil #1 \rceil}
\newcommand{\flo}[1]{\lfloor #1 \rfloor}

\begin{document}

\title{Influence and Dynamic Behavior in Random Boolean Networks}

\author{C. Seshadhri}
\author{Yevgeniy Vorobeychik}
\author{Jackson R. Mayo}
\author{Robert C. Armstrong}
\author{Joseph R. Ruthruff}
\affiliation{Sandia National Laboratories, P.O. Box 969, Livermore, California 94551-0969, USA}

\begin{abstract}
We present a rigorous mathematical framework for
analyzing dynamics of a broad class of Boolean network models.
We use this framework to provide the first formal proof of many of the standard critical
transition results in Boolean network analysis, and offer analogous
characterizations for novel classes of random Boolean networks.
We precisely connect the short-run dynamic behavior
of a Boolean network to the average influence of the transfer functions.
We show that some of the assumptions traditionally made in the
more common mean-field analysis of Boolean networks do not hold in
general.
For example, we offer some evidence that imbalance, or expected internal
inhomogeneity, of transfer functions is a crucial feature that tends to
drive quiescent behavior far more strongly than previously observed.
\end{abstract}

\maketitle

\paragraph*{Introduction.}

Complex systems can usually be represented as a network of
interdependent functional units.
Boolean networks were proposed by Kauffman as models of
genetic regulatory networks~\cite{Kauffman69,Kauffman93} and have
received considerable attention across several scientific disciplines.
They model a variety of complex phenomena,
particularly in theoretical biology and
physics~\cite{Harris02,Shmulevich04,Moreira05,Aldana03,Kauffman03,Shmulevich03}.

A Boolean network $\cN$ with $n$ nodes can be described by a directed graph $G= (V,E)$
and a set of \emph{transfer functions}.
We use $V$ and $E$ to denote the sets of nodes and edges respectively, and
denote the indegree of node $i$ by $K_i$.
Each node $i$ is assigned a $K_i$-ary Boolean function
$f_i:\{-1,+1\}^{K_i} \rightarrow \{-1,+1\}$, termed \emph{transfer
  function}.
If the state of node $i$ at time $t$ is $x_i(t)$, its state at time
$t+1$ is described by
\[
x_i(t+1) = f_i(x_{i_1}(t),\ldots,x_{i_{K_i}}(t)).
\]
\noindent
The state of $\cN$ at time $t$ is just the vector $(x_1(t), x_2(t), \ldots, x_n(t))$.

Boolean networks are studied by positing a distribution of
graph topologies and Boolean functions from which independent random draws are made.
We denote the distribution of transfer functions by $\cT$. An early observation was that when the indegree of a network
is fixed at $K$ and each transfer function is chosen uniformly
randomly from the set of all $K$-input possibilities,
the network dynamics undergo a critical transition at $K=2$, such that
for $K < 2$ the network behavior is quiescent and small perturbations
die out, while for $K > 2$ it exhibits chaotic features~\cite{Kauffman93}.
This result has been generalized to non-homogeneous distributions of
transfer functions, when the output bit is set to 1 with
probability $p$ (called \emph{bias}) independently for every possible
input string~\cite{Derrida86}. The resulting critical boundary
is described by the equation $2Kp(1-p) = 1$.

All analysis of Boolean networks to date uses mean-field
approximations, an annealed
approximation~\cite{Derrida86}, simulation
studies~\cite{Kauffman69,Kauffman03}, or combinations of these, to
understand the dynamic behavior. 
Many previous studies rely solely on 
short-run characteristics (e.g., Derrida plots that
consider only a very short, often only a single-step, horizon~\cite{Kauffman03,Shmulevich04,Moreira05}) and
extrapolate to understand long-term dynamics.
Hamming distance between Boolean network states that
diverges exponentially over time for small perturbations to initial
state suggests sensitivity to initial
conditions typically associated with chaotic dynamical
systems. Nonetheless, the connection
between short-run and long-run sensitivity is not a foregone
conclusion~\cite{Ghanbarnejad11} and remains an open question.

We provide a formal mathematical framework to analyze the behavior of Booleam networks over a logarithmic (in the size of the graph) number of discrete
time steps, and give conditions for exponential divergence in Hamming
distance in terms of the indegree distribution and
influence of transfer functions in $\cT$.

\paragraph*{Assumptions.}

We assume that the Boolean network $\cN$ is constructed as follows.
First, we specify an indegree distribution $\cD$ with a maximum possible
indegree $K_{max}$, and for each node $i$ independently draw its
indegree $K_i \sim \cD$.
We then construct $G$ by choosing each of the $K_i$ neighbors of every node $i$
uniformly at random from all $n$ nodes.
Next, for each node $i$ we independently choose a $K_i$-input transfer
function according to $\cT$.
We assume that the family $\cT$ has \emph{either} of the following properties:
\begin{itemize}
	\item {\bf Full independence}: Each entry in the truth table
          of a transfer function is i.i.d., \emph{or}
	\item {\bf Balanced on average}: 
          Transfer functions drawn from $\cT$ have, on average, an
          equal number of $+1$ and $-1$ output entries in the truth
          table. Formally, $\pr_{f,x}[f(x) = +1]$ $= 1/2$, where $\pr_{f,x}$ denotes the probability of an event when $f$ is drawn
from $\cT$, and input $x$ for $f$ is chosen uniformly at random.
\end{itemize}
\paragraph*{Influence.}

The notion of \emph{influence} of variables on
Boolean functions was defined by Kahn \emph{et al.}~\cite{Kahn88} and introduced to the study of Boolean networks by Shmulevich and
Kauffman~\cite{Shmulevich04}.
The \emph{influence} of input $i$ on a Boolean function $f$, denoted
by $\Inf_i(f)$, is
\[
\Inf_i(f) = \pr_x[f(x) \neq f(\flip{x}{i})],
\]
\noindent
where $\flip{x}{i}$ is the same as $x$ in all coordinates except $i$.
Given a distribution $\cT$ of transfer functions, let $\cT_d$
denote the induced distribution over $d$-input transfer functions.
The expected total influence under $\cT_d$, denoted by $\I(\cT_d)$, is
$\EX_{f\sim \cT_d}[\sum_i \Inf_i(f)]$.
When $\cT_d$ is clear from the context we write this
simply as $I(d)$.
Suppose that we have an indegree distribution where $p(d)$
is the probability that indegree is $d$.
We show that the quantity that characterizes the
dynamic behavior of Boolean networks is 
\[
\cI = \sum_{d=1}^{K_{max}}p(d)\I(d).
\]

\paragraph*{Main Result.}

We present our main result that characterizes
dynamic behavior of Boolean networks under the assumptions stated above.
Define $t^* = \log n/(4\log K_{max})$.
The following theorem tracks the evolution of Hamming distance up to time $t^*$,
starting with a small (single-bit) perturbation. 
We note that our theorem applies for any distribution of indegrees
with a maximum bounded by $K_{max}$, though increasing density
($K_{max}$) shortens the effective horizon $t^*$.
\begin{theorem} \label{thm:avg-inf} Choose a random Boolean network
  $\cN$ having a random graph $G$ with $n$ nodes
and a distribution of transfer functions $\cT$. Evolve $\cN$ in
parallel from a uniform random starting state
$x$ and its flip perturbation $x^{(i)}$ (with a uniform random $i$).
The expected Hamming distance between the respective states of $\cN$ at time $t \leq t^*$
lies in the range $\cI^t \pm 1/n^{1/4}$.
\end{theorem}
The proof of this theorem is non-trivial and is provided in the
supplement. 
It shows that the effects of flip perturbations vanish when $\cI < 1$
while perturbations diverge exponentially when $\cI > 1$.
Thus, criticality of the system is equivalent to $\cI = 1$.

Much of the past work assumed (or explicitly
stated) that it suffices to consider the
expected influence value $I(K)$ for the \emph{mean} indegree $K$.
A direct consequence of Theorem~\ref{thm:avg-inf} is
that $I(K)$ characterizes a critical transition \emph{iff $I(d)$ is affine}.
To see this, observe that $I(K) = \cI$ iff
\(
I(K) = I\left(\sum_d dp(d)\right) = \sum_d p(d)I(d).
\)
This is true if and only if $I(d)$ is affine.

\paragraph*{Applications.}
In this section we use Theorem~\ref{thm:avg-inf} to recover most of the characterizations of
critical indegree thresholds to date and prove results for new natural classes of transfer functions.
We show that
our assumptions are crucial in obtaining the observed results.
An important step in applying the theorem is computing the quantity
$I(d)$ for a given class of transfer functions $\cT$.
The following proposition (proven in the online supplement)
facilitates this process.
Let $\cB^d$ denote a $d$-dimensional Boolean hypercube.
The edges of $\cB^d$ connect pairs of elements with Hamming distance $1$.
A function $f:\cB^d \rightarrow \cB$ can be represented by labeling
element $x \in \cB^d$ by $f(x)$.
An edge of $\cB^d$ is called \emph{$f$-bichromatic}
if one endpoint is labeled $+1$ and the other $-1$.
%
\begin{prop} \label{prop:bool}
Consider a distribution $\cT_d$ over $d$-input functions.
Then
\[
\I(\cT_d) = \frac{\EX_{f \sim \cT_d} [\textrm{\# $f$-bichromatic edges}]}{2^{d-1}}.
\]
\end{prop}

{\bf Uniform random transfer functions.} We begin with the classical model of random Boolean networks in which
each entry in the truth table of a transfer function is chosen to be $+1$ and
$-1$ with equal probability.
It has previously been observed that the critical transition occurs at
mean indegree $K=2$~\cite{Derrida86}.
We now demonstrate that it is a simple corollary of our theorem.
First, we need to compute $I(d)$ using Proposition~\ref{prop:bool}.
In this model, the probability that an edge is $f$-bichromatic is exactly $1/2$. 
Hence, $\I(d) = (\textrm{total number of edges})/2^d$.
Since the total number of edges (of $\cB^d$) is $d2^{d-1}$, we obtain $\I(d) =
d/2$.
Notice that $\I(d)$ is linear in this case, and, consequently,
considering $\I(K) = K/2$ suffices for any distribution $p(d)$.
Applying Theorem~\ref{thm:avg-inf}  then gives us the well-known critical
transition at $K = 2$.

{\bf Transfer functions with a bias $p$.} A simple generalization of
uniform random transfer functions is to introduce a bias, that is, a
probability $p$ that an entry in the truth table is $+1$ (but still
filling in the truth table with i.i.d. entries)~\cite{Kauffman93}.
In this case, the probability that an edge is $f$-bichromatic is $2p(1-p)$
and therefore $\I(d) = 2dp(1-p)$.
Since $I(d)$ is linear, we can characterize the critical transition in this case at
$2Kp(1-p) = 1$ for any indegree distribution with mean $K$.

{\bf Canalizing functions.} Kauffman~\cite{Kauffman93} and others have
observed that since uniform random transfer functions are typically
chaotic, they are unlikely to represent a distribution of
transfer functions that accurately models real phenomena, such as
genetic regulatory networks.
Biased transfer functions only partially resolve
this, as they still tend to fall easily into a chaotic regime for a
rather broad range of $p$~\cite{Aldana03}.
Empirical studies of genetic networks suggest another
class of transfer functions called \emph{canalizing}.
A canalizing function has at least one input, $i$, such that
there is some value of that input, $v_i$, that determines the value of
the Boolean function.
Shmulevich and Kauffman~\cite{Shmulevich04} show heuristically that canalizing functions have
$\I(K) = (K+1)/4$ and thus exhibit a critical transition at $K = 3$.
We now show that this is a corollary of our theorem,
using Proposition~\ref{prop:bool} to obtain $I(d)$.

To compute $I(d)$, fix (without loss of generality) the canalizing
input index to be $1$ and the canalizing input and output values
to $+1$. Consider the distribution of functions
conditional on these properties. By symmetry, the expected
number of bichromatic edges conditional on this is the same
as the overall expectation. Hence, we can focus on choosing $f$
from this conditional distribution.
Split the hypercube $\cB^d$ into the $(d-1)$-dimensional
sub-hypercubes $\cB'$ and $\cB''$ such that $\cB'$ has all inputs with $x_1 = +1$ and $\cB''$ has all
inputs that have $x_1 = -1$. 
Edges can be partitioned into three groups
$E', E'', E^*$. The set of edges $E'$ (resp.\ $E''$) are those that are internal
to $\cB'$ (resp.\ $\cB''$). The set of edges $E^*$ have endpoints
in both $\cB$ and $\cB'$. Note that $|E'| = |E''| = (d-1)2^{d-2}$, and
$|E^*| = 2^{d-1}$. Because the function is canalizing, the edges in
$E'$ are all $f$-monochromatic, and all other edges are $f$-bichromatic with probability $1/2$. Hence,
the expected number of bichromatic edges is
\(
((d-1)2^{d-2} + 2^{d-1})/2 = 2^{d-1} (d+1)/4.
\)
By Proposition~\ref{prop:bool}, we then have $\I(d) = (d+1)/4$.
Since this is affine in $d$, we can conclude that $I(K) = (K+1)/4$
characterizes the short-run dynamic behavior for any indegree
distribution with mean $K$.

{\bf Threshold functions.} A threshold function $f(x)$ with $d$ inputs has the form
$\mathrm{sgn}[f^*(x)]$ with
\[
f^*(x) = \frac{1}{d}\sum_{i \le d} w_i x_i - \theta,
\]
\noindent
where $x_i$ is the value of input $i$, $w_i \in \{-1,+1\}$ is its
weight, which has a natural interpretation of an input being
inhibiting ($w_i = -1$) or excitatory ($w_i = +1$) in regulatory
networks, and $\theta$ is a real number in $[-1,+1]$ representing an
inhibiting/excitatory threshold for $f$.
Such 2-input threshold functions have
been studied by Greil and Drossel~\cite{Greil07} and Szejka et al.~\cite{Szejka08} and are classified
as biologically meaningful by Raeymaekers~\cite{Raeymaekers02}.
We now use Theorem~\ref{thm:avg-inf} to show that random threshold functions lead
to criticality for any indegree distribution.

Consider $\cT$ in which the value of $w_i$ for each input $i$, as
well as $\theta$, are chosen uniformly at random.
To compute $\I(d)$, consider a threshold function with threshold
$\theta$ and an edge $(x,\flip{x}{i})$.
This edge is bichromatic exactly when the $\theta$ lies between
$f(x)$ and $f(\flip{x}{i})$. Note that $|f^*(x) - f^*(\flip{x}{i})| = 2/d$, regardless
of the values $w_1,\ldots,w_d$.
Since the range of $\theta$ has size $2$, the probability that this happens is $(2/d)/2 = 1/d$. So
$\I(d) = (\textrm{\# of edges})/d2^{d-1} = 1$.
Since it is independent of $d$, the result follows immediately by Theorem~\ref{thm:avg-inf}.

{\bf Majority function.} An important specific threshold function is a
majority function, which has $w_i = 1$ for all inputs $i$ and $\theta
= 0$.
Suppose $\cT$ consists exclusively of majority functions.
We demonstrate that the quiescence-chaos transition properties of this
class are very different from those of general threshold functions.
One detail that needs to be specified for $\cT$ is what to do when the
number of positive and negative inputs is exactly balanced.
To satisfy the condition that $\cT$ is balanced in expectation, we let
the output be $+1$ or $-1$ with equal probability in such an instance
(for a specific majority function this choice is determined, but it is
randomized for any majority function generated from $\cT$).
Given this $\cT$, we now show that
\[
I(d) = \frac{\cei{d/2}}{2^{d-1}} \binom{d}{\flo{d/2}}.
\]
\noindent
When $d$ is odd, bichromatic edges are 
those that connect the $\flo{d/2}$-level to the $\cei{d/2}$-level.
For $d$ even, these are the edges connecting the $d/2$-level
to the $(d/2-1)$-level (or the $(d/2+1)$-level). In either case,
the number of these edges is $\cei{d/2}\binom{d}{\flo{d/2}}$, giving
$I(d)$ as above.
Consequently, when $d = 1$ or $2$, $I(d) = 1$, while for $d \geq
3$, $I(d) \geq 3/2$.
Thus, if a Boolean network has a fixed indegree $K$, it is critical
for $K \leq 2$ and chaotic for $K > 2$.

\begin{figure}
  \centering
  \includegraphics[width=.3\textwidth,trim=0 0 0 5,clip]{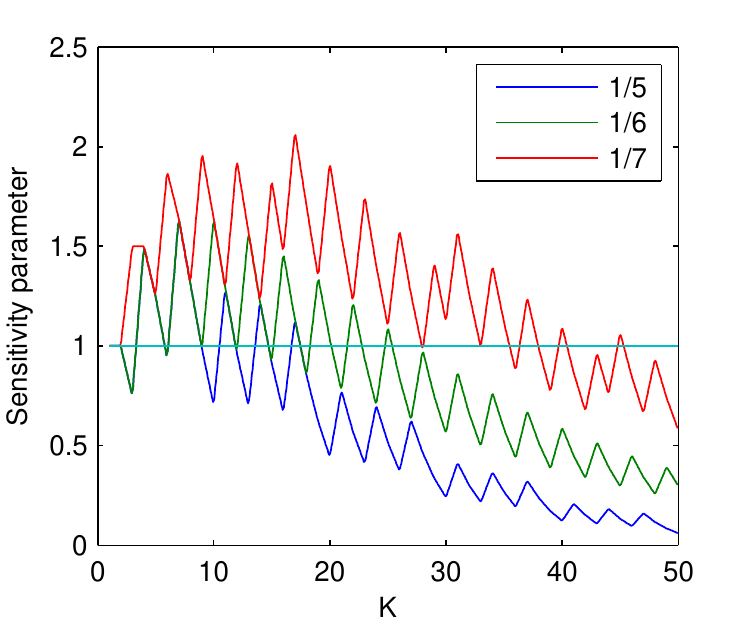}
  \caption{The $x$-axis is $K$. The $y$-axis gives the average
    influence parameter $I(K)$. 
  We show the cases where $\rho = 1/5, 1/6, 1/7$. For larger $\rho$ we reach quiescent
  behavior more rapidly with 	increasing $K$.}
  \label{fig:str-maj}
\end{figure}

{\bf Strong majority function.} We now show an interesting and natural class of functions
where the expected average influence goes \emph{down} as the indegree
$d$ increases.
Consider threshold functions where $w_i = 1$ for all inputs $i$ and the
threshold is either $\theta$ or $-\theta$ with equal probability for
some fixed $\theta \in [0,1]$.
For example, when $\theta = 1/3$, the function returns $+1$ iff
a 2/3 majority of inputs have value $+1$.
For this class of functions, bichromatic edges are those that connect
the $\flo{d/2 + \rho d}$-level to the $\cei{d/2 + \rho d}$-level,
where $\rho = \theta/2$.
Thus, the expected number of bichromatic edges for a fixed $d$ is
\[
B_e = (d - \flo{d/2 - \rho d}) \binom{d}{\flo{d/2 + \rho d}},
\]
\noindent
and, consequently,
\(
I(d) = B_e/2^{d-1}.
\)
In Figure~\ref{fig:str-maj} we plot $I(K)$, where $K$ is a fixed
indegree, for different values of $\rho$.
There are two rather remarkable observations to be made about this
class of transfer functions:
first, the sawtooth behavior of $I(K)$, and second, that the Boolean
network actually becomes \emph{more quiescent} with increasing $K$.
To our knowledge, this is the first example in which there is no
single critical transition from order to chaos, and increasing
connectivity leads to greater order.
We show that for $d$ large enough, $I(d)$ tends to 0.
For convenience, assume that $d$ is an even integer and $\theta d$
is non-integral.
By tail bounds on binomial coefficients, $2^{-d}\sum_{r \geq \flo{d/2 + \rho d}} \binom{d}{r} < 2^{-cd}$
for some constant $c$. (This can be proven using a Chernoff bound, such as
Theorem 4.1 in~\cite{MR}.) Hence $\I(d)< 1$ for large enough $d$, and 
tends to zero as $d$ increases.
We had previously noted that it is commonly assumed that
$I(d)$ is linear in $d$.
Strong majority transfer functions feature $I(d)$ that is clearly
non-linear, and we therefore expect this assumption to be
consequential.
To illustrate, consider two network structures: one with a
fixed $K=4$, and another where the indegree distribution follows a
power law with mean $K=4$.
Using $\theta = 1/3$, in the former, we get $\cI = I(K) = 1.5$, while in the latter (with $K_{max}
= 100$), $\cI = 0.79$.
Thus, while a fixed $K$ yields decidedly chaotic dynamics, using a
power law distribution with the same mean indegree produces
quiescence.
\begin{figure*}[ht]
\centering
\includegraphics[width=7in]{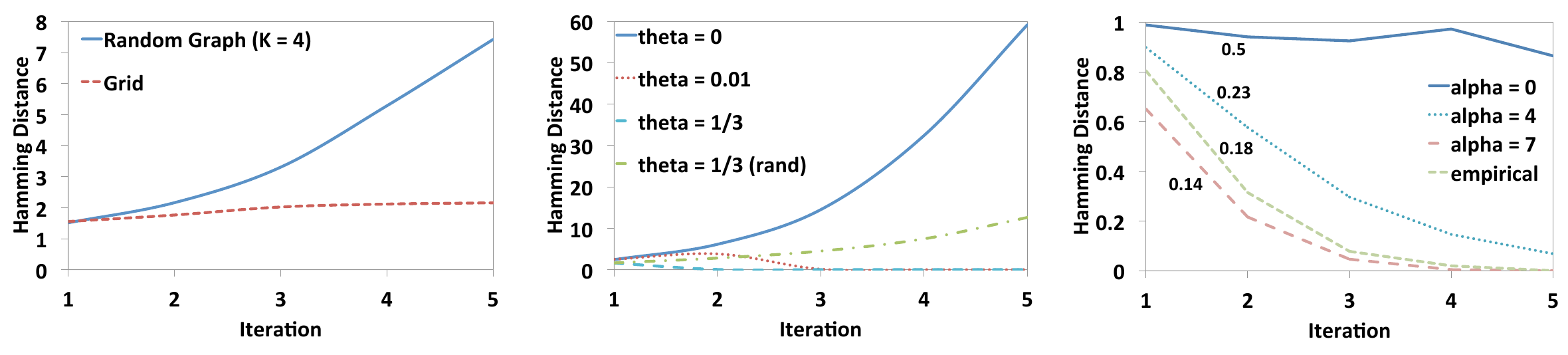}
\caption{Boolean network dynamics (Hamming distance over time,
  starting with single-bit perturbations).  
  Left: comparing random graph with $K=4$ and a grid.
  Middle: random networks with fixed $K=10$ and
  \emph{unbalanced} strong majorities.
  Right: random networks with $K=5$, using
  nested canalizing functions, with degree of imbalance increasing
  with $\alpha$, and the empirical distribution
  of 5-input transfer functions based on yeast regulatory
  networks~\cite{Kauffman03,Harris02}.  
}
\label{F:balance}
\end{figure*}

{\bf The importance of graph structure.} Our results rely fundamentally on the fact that the
inputs into each node are chosen independently.
The fact that the size of the neighborhood at distance $t$ grows exponentially
with $t$ is crucial for our proofs.
Furthermore (for the random graphs we sample from), this neighborhood
is a root directed tree, when $t < t^*$.
When graphs exhibit only polynomial local growth, we do
not expect chaotic dynamic behavior even when other conditions for
it are met.
We illustrate this point in Figure~\ref{F:balance} (left), which compares a random network with
$K=4$ to a grid (a bidirectional square lattice that also has $K=4$).
While both initially appear to be in a chaotic regime, the Hamming
distance stops diverging for a grid, but diverges exponentially in the random network.
\leaveout{
\begin{figure}
\centering
\includegraphics[width=.3\textwidth]{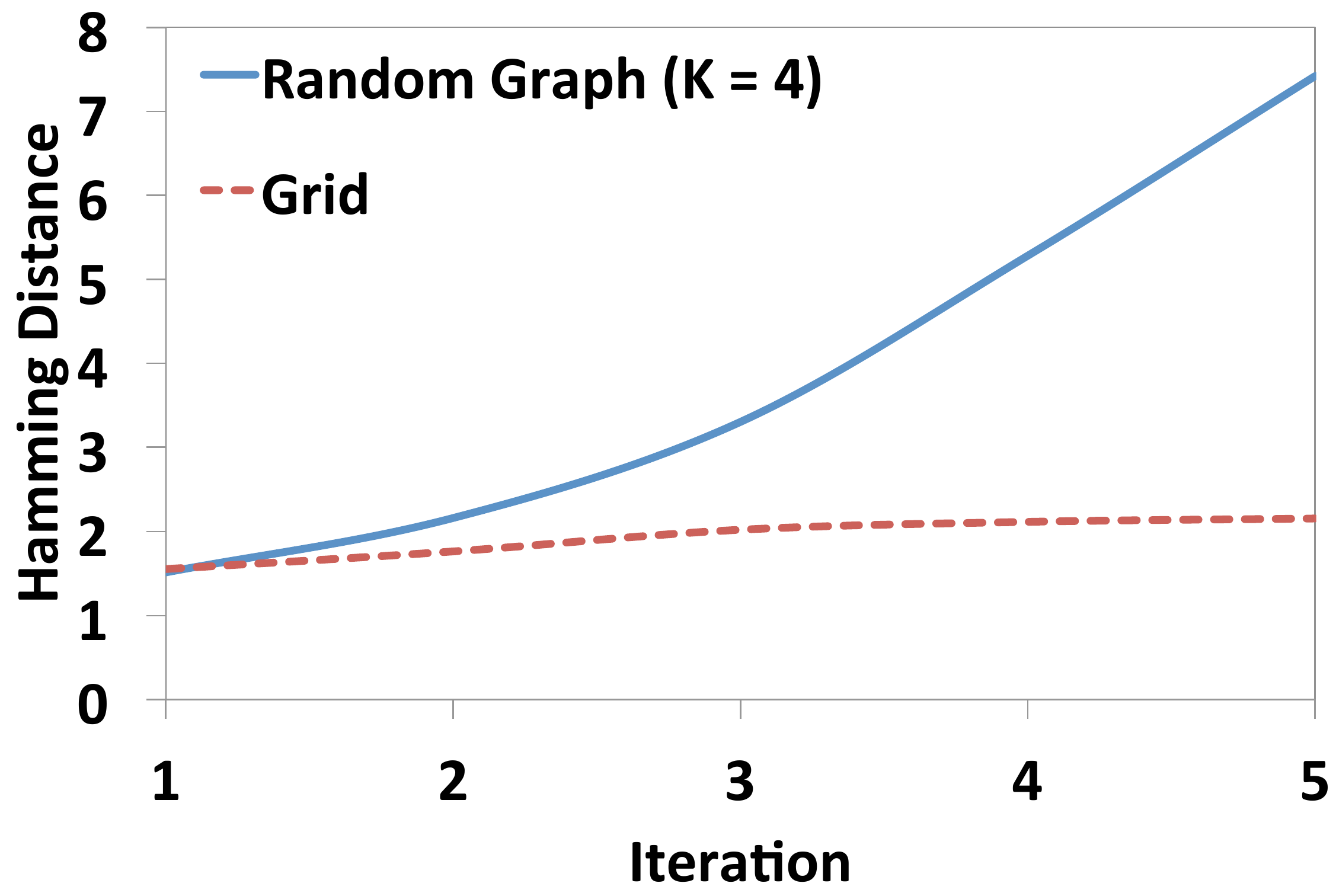}
\caption{Expected Hamming distance over time after a single-bit
  perturbation for a random graph with $K=4$ and a grid.}
\label{F:graph}
\end{figure}
}

{\bf The importance of being balanced.} The assumption that 
$\cT$ is balanced is crucial.
Balance has previously been noted to play an important role in
determining the order to chaos transition, but entirely under the
assumption that each truth table entry is i.i.d.~\cite{Kauffman93}.
It has been pointed out that much of the resulting space of
parameter values gives rise to chaotic dynamics~\cite{Aldana03}.
What we now demonstrate is that this observation is largely an
artifact of independence, and when truth table entries are not
independently distributed, even a slight deviation from balance
(homogeneity) may push Boolean network dynamics to quiescence.
Consider networks in which every transfer function is a strong
majority (with $\theta = 0$ being a simple majority).
We get a balanced distribution of transfer function by choosing between $\theta$ and $-\theta$
with equal probability. An \emph{imbalanced distribution} is obtained
by choosing only one of them.
Figure~\ref{F:balance} (middle) shows several examples of how the
Hamming distance evolves for different values of $\theta$, and
contrasts the balanced and unbalanced settings.
The difference could hardly be more dramatic: even a slight deviation
from simple majority ($\theta = 0.01$) is a difference between chaos
and quiescence; indeed, it is instructive to see the initial increase
in Hamming distance for the imbalanced strong majority with
$\theta=0.01$, only to be ultimately suppressed.
Similarly, we can compare the balanced and unbalanced versions of strong
majorities with $\theta = 1/3$: the balanced version is clearly
chaotic, while in the network with the unbalanced analogue, initial
perturbation effects erode within two iterations.
A similar picture emerges when we consider nested canalizing
functions, previously offered as an explanation of robustness in
genetic regulatory networks~\cite{Harris02,Kauffman03}.
Classes of these are generated by a parameter $\alpha$ that
governs the fraction of 1's in the transfer function
truth table, with larger values of $\alpha$ leading to greater
imbalance.
Figure~\ref{F:balance} (right) compares evolution of networks with
nested canalizing functions, as well as with transfer functions
following an empirical distribution of transfer functions based on
regulatory networks~\cite{Kauffman03}.
We see that the main driver of quiescence appears to be the
internal inhomogeneity of transfer functions, rather than canalizing
properties.
\leaveout{
Figure~\ref{F:balance} (right) makes a similar comparison, but using
a part of an actual yeast gene regulatory network reported in Kauffman
et al.~\cite{Kauffman03}.
In both cases, the main driver of quiescence appears to be the lack of
internal homogeneity of transfer functions, rather than canalizing
properties.
Indeed, looking at the empirically derived regulatory network in
Figure~\ref{F:balance} right, initial influence is above 3, suggesting
strongly chaotic dynamics, only to be squashed to below 1 within
several iterations.
}

{\it Sandia is a multiprogram laboratory operated by Sandia Corporation, a wholly owned subsidiary of Lockheed Martin Corporation, for the U.S. Department of Energy under contract DE-AC04-94AL85000.}


\bibliographystyle{apsrev4-1}
%

\newpage

\appendix

\centerline{\Large \bf Supplemental Material}
\bigskip

The following supplemental material for the paper includes all formalization and
proof details.

\section{Preliminaries and notation} \label{sec:prelims}

We will be using probabilities very heavily, so it will help to set some
notations. Capital letters $X,Y,Z$ will used
to denote random variables. Events are denoted by calligraphic letters
$\cE,\cF,\ldots$. The probability of an event $\cE$ is denoted by $\pr(\cE)$.
The expectation of a random variable $X$ is denoted by $\EX[X]$.
The random variable $X$ conditioned on event $\cE$ is denoted by $X | \cE$.

\subsection{Preliminaries} \label{sec:bool}

{\bf Graphs:} We will always deal with directed graphs $G = (V,E)$.
The set $V$ will be $[n]$, the set of positive integers upto $n$.
For every $i \in V$, $N^+(i)$ (resp. $N^-(i)$) denotes the set of
out (resp. in) neighbors of $i$ in $G$. We use $d^+(i)$ to denote the outdegree
of $i$ (similarly, define $d^-(i)$).\\\\
{\bf Boolean strings and functions:} We will use the set $\{-1,+1\}$ (instead of $\{0,1\}$) to denote
bits, thereby aligning ourselves with the theory of Boolean functions.
The $n$-dimensional Boolean hypercube is denoted by $\cB^n = \{-1,+1\}^n$.
This is also a representation for all $n$ bit strings.
A \emph{Boolean function} is a function $f:\cB^n \rightarrow \cB^1$.

For any two elements $x ,y \in \cB_n$, $\Delta(x,y)$
is the Hamming distance between $x$ and $y$. For $x \in \cB^n$,
$\flip{x}{i}$ is the unique element of $\cB^n$ which is the same
as $x$ on all coordinates except for the $i$th coordinate.
We use $x_i$ to denote the $i$th coordinate of $x$.

We use $\pr_x(\cE)$ (where $\cE$ is some event) to denote
the probability of $\cE$ over the uniform distribution
of $x$ (which means we choose an $x$ uniformly at random
from $\cB^n$ and check the probability of $\cE$).
Similarly, $\EX_x[\ldots]$ denotes the expectation over a random
uniform $x$. \\\\
{\bf Boolean networks:} A Boolean network $\cN$ consists of a directed graph
$G$ and a set of transfer functions $T$. The set $T$ has
a transfer function $\tau_v: \cB^{d^-(i)} \longrightarrow \{-1,+1\}$,
for each $i \in V$.

The state of a Boolean network is just an assignment of $\{-1,+1\}$
to each vertex in $V$. This can be represented as an $n$-bit
string, or alternately, an element $x$ of the $n$-dimensional Boolean hypercube
$\cB^n = \{-1,+1\}^n$. 

Suppose $\cN$ starts at the state $x$.
The state at vertex $i$ after $t$ steps of $\cN$ is denoted
by the Boolean function $f_{t,i}(x)$. The function $f_t(x): \cB^n \rightarrow \cB^n$
is state of $\cN$ after $t$-steps (so this is an $n$-dimensional vector
having $f_{t,i}(x)$ as its $i$th coordinate).\\\\
{\bf Assumptions on Boolean network:} We will analyze Boolean networks that
arise from a particular distribution. First, the graph $G$
is chosen from a random distribution. We will assume some fixed indegree distribution $\cD$. 
(This is simply a distribution on positive integers.) 
For each vertex $i$, we \emph{independently} choose $d^-(i)$ from $\cD$. Then,
we choose $d^-(i)$ uniform random vertices (without replacement) to be the in-neighborhood ($N^-(i)$)
of $i$. We denote the average indegree by $K$, and the maximum possible
indegree by $K_{max}$.

Next, we assume that there is a distribution $\cT$ on transfer functions.
Formally, this is a union of distributions $\cT_d$, where this family
only contains Boolean functions that take $d$ inputs. For each vertex $i$,
we first choose an independent function $\tau_i(x_1,x_2,\ldots,x_{d^-(i)})$ 
from $\cT_{d^-(i)}$. Sort $N^-(i)$ (according to its label) to get
$v_1, v_2, \ldots, v_{d^-(i)}$. Assign the vertex $v_j$ to input $x_j$
of $\tau_i$. This gives us the transfer function for vertex $i$.

We will assume that the family $\cT$ has \emph{either} of the following properties:

\begin{itemize}
	\item Full independence: A random function in $\cT$ is generated by taking an 
	empty truth table, and filling in each entry independently with the same distribution.
	\item Balance on average: A uniform random member of $\cT$ evaluated on a uniform random
	input outputs $+1$ with probability $1/2$. Formally, $\pr_{x,\tau}[\tau(x) = +1]$ $= 1/2$
	$= \pr_{x,\tau}[\tau(x) = -1]$
\end{itemize}

\subsection{Influences}

We now discuss some of our main definitions. The following is one of the most
important concepts.

\begin{definition} \label{def:inf} For a Boolean function $f$ and coordinate $i$,
the \emph{influence} of $i$ on $f$, denoted by $\Inf_i(f)$ is $\pr_x[f(x) \neq f(\flip{x}{i})]$.
The \emph{total influence} of $f$ is $\sum_i \Inf_i(f)$ and the \emph{average influence}
of $f$ is $\frac{1}{n} \sum_i \Inf_i(f)$.

For a distribution $\cT = \cup_d \cT_d$ of transfer functions, the influence
of the distribution $\cT_d$ is denoted by $\I(\cT_d)$. Formally, $\I(\cT_d) = \EX_\tau[\sum_i \Inf_i(\tau)]$.
Often, when the defintion of $\cT$ is unambiguous, we write this as $\I_d$.
\end{definition}

\begin{definition} \label{def:inf-net} The \emph{influence of $i$ at time $t$} on $\cN$,
denoted by $\Inf_{t,i}(\cN)$,
is $\EX_x[\Delta(f_t(x),f_t(\flip{x}{i}))]$. The \emph{average influence at time $t$} 
is $\frac{1}{n} \sum_i \Inf_{t,i}(\cN)$.
\end{definition}

\begin{claim} \label{clm:avg-inf} The average influence of $\cN$ at time $t$ can
be expressed as $\frac{1}{n} \sum_{1 \leq i,j \leq n} \Inf_i(f_{t,j})$.
\end{claim}

\begin{proof} Let us focus on $\Inf_i(\cN)$. We choose a uniform random $x$ and 
evolve $\cN$ from the states $x$ and $\flip{x}{i}$. Let $\chi(t,j)$ be the indicator
random variable for the event that $f_{t,j}(x) \neq f_{t,j}(\flip{x}{i})$. Note
that $\EX_x[\chi(t,j)] = \Inf_i(f_{t,j})$. By the definition of Hamming distance
and linearity of expectation,
\begin{eqnarray*}
	\Inf_i(\cN) & = & \EX_x[\Delta(f(x),f(\flip{x}{i}))] \\
	& = & \EX_x[\sum_j \chi(t,j)] = \sum_j \EX_x[\chi(t,j)] = \sum_j \Inf_i(f_{t,j})
\end{eqnarray*}
Averaging this equality over all $i$ completes the proof.
\qed\\
\end{proof}

We will need the following simple facts about influences. (This is a restatement
of Proposition~\ref{prop:bool}.

\begin{prop} \label{prop:bool2}
\begin{itemize}
	\item Consider a function $f: \cB^d \rightarrow \cB$.
An edge of the Boolean hypercube $\cB^d$ is called \emph{bichromatic}
if one endpoint is labelled $+1$ and the other is labelled $-1$.
Then $\sum_{i \leq d} \Inf_i(f) =$ $(\textrm{\# bichromatic edges})/2^{d-1}$.

Consider a distribution $\cT$ over functions $f:\cB^d \rightarrow \cB$.
$$ \EX[\sum_{i \leq d} \Inf_i(f)] = (1/2^{d-1}) \sum_{e: \textrm{edge} \ \in \cB^d} \Pr[\textrm{$e$ is bichromatic}] $$

	\item  For any Boolean function $f$ and input index $i$, 
\begin{eqnarray*}
 & & \Pr_x[(f(x) = 1) \wedge (f(x) \neq f(\flip{x}{i}))] \\
 & = & \Pr_x[(f(x) = -1) \wedge (f(x) \neq f(\flip{x}{i}))] 
= \Inf_i(f)/2 
\end{eqnarray*}
\end{itemize}
\end{prop}

\begin{proof} Consider all pairs $(y,\flip{y}{i})$, where the $i$th bit of $y$ is $1$. These
pairs form a partition of the hypercube and are actually all edges of the hypercube parallel
to the $i$th dimension. The influence $\Inf_i(f)$ is exactly the probability that a uniformly random $x$
belongs to a bichromatic pair. Let $B_i$ be the number of bichromatic pairs.
Noting that the total number of edges parallel to the $i$th
dimension is $2^{d-1}$, $\Inf_i(f) = B_i/2^{d-1}$. Summing over all $i$,
we get that $\sum_{i \leq d} \Inf_i(f) =$ $(\textrm{\# bichromatic edges})/2^{d-1}$.
To deal with a distribution, we simply apply linear of expectation to
this bound.

Now for the second part. When does the event $(f(x) = 1) \wedge (f(x) \neq f(\flip{x}{i}))$
happen? This happens when $x$ belongs to a bichromatic pair, and $f(x) = 1$. Exactly
half the member of bichromatic pairs have value $1$ (or $-1$). Hence, 
$\Pr_x[(f(x) = 1) \wedge (f(x) \neq f(\flip{x}{i}))] = \Inf_i(f)/2$.
\end{proof}

We restate Theorem~\ref{thm:avg-inf} for convenience.

\begin{thm} \label{thm:avg-inf2} The average influence at time $t \leq t^*$
for the Boolean network $\cN$ 
lies in the range $(\sum_{d \geq 1} p_d\I_d)^t \pm 1/n^{1/4}$.
\end{thm}

\section{Proof strategy} \label{sec:strat}

Before diving into the gory details of the proof of our main result,
we offer a high level intuition of the overall proof. 
The proof essentially consists of two central steps,
Lemma~\ref{lem:tree} and Claim~\ref{clm:g-tree}, discussed in Sections~\ref{sec:trees}
and~\ref{sec:top} respectively. 
Lemma~\ref{lem:tree} considers
an idealized situation where the underlying graph of a Boolean network is simply
a root-directed tree. Any modification of the state of a leaf travels
up the tree, possibly
affecting the root. In this case, we can give an exact expression of the influence of leaves
on the root. 
Our key next step is Claim~\ref{clm:g-tree}, which proves that for our
distribution of graph topologies (which is closely related to the configuration model),
the logarithmic-distance neighborhood of most nodes looks like a root-directed tree. Note that
this does not imply that the graph itself decomposes into disjoint
tree, since the trees rooted at each node are interconnected in complicated ways. 

These two steps are combined together for the final proof in
Section~\ref{sec:final}. This proof makes heavy use of the linearity of expectation
and some conditional probability arguments.  It enables us to perform exact short-run analysis of the Boolean
network by only considering local neighborhoods of an ``average" node.

At the core of the proof of Lemma~\ref{lem:tree} is a straightforward induction
argument. Consider a tree network, where we change the state at some
leaf. 
The catch is that induction requires
the family of transfer functions to satisfy the technical conditions of balance or full independence.
This is one of the major insights of this work, since
these conditions on transfer function families have generally been implicit in previous results. The proof forces us to 
make these conditions explicit. Section~\ref{sec:trees} has the details.

The proof of Claim~\ref{clm:g-tree} consists of a combinatorial probability calculation. We are generating
our graph through the randomized process of choosing the (immediate) neighborhood for each node independently (and uniformly)
at random. We show that the probability that the short-distance neighborhood of a node contains a cycle
is extremely small. 
The formal proof is given in Section~\ref{sec:top}.

\section{Influences on trees} \label{sec:trees}

In this section, we will make some arguments about tree networks. 
Let $T$ be
a directed tree (so all edges point towards the root) with all leaves
at the same depth $h$. We are interested in the influence of the leaf
variables on the root $r$. Let the function giving the state of the 
root $r$ at time $t$ be $f_{t,r}$. Note that $f_{h,r}$ is \emph{only}
a function of the leaves, since there is no feedback in this graph.
Note that we are not particularly bothered with what happens in the leaves
are time step $1$ (since those values are not even defined). We are merely
interested in how the values at the leaves will propagate up the tree.
For each $v$, the transfer function $\tau_v$ is chosen from $\cT$ (technically, from $\cT_{d^-(v)}$).

\begin{claim} \label{clm:tree} Let $T$ be a tree of depth $h$.
For a leaf $\ell$, let the path to the root be
$v_0 = \ell, v_1, v_2, \ldots, v_h = r$.
Then, $\EX_\cT[\Inf_\ell(f_{h,r})] = \prod_{i = 1}^{h} \EX_{\cT}[\Inf_{v_{i-1}}(\tau_{v_{i}})]$.
(We remind the reader that $\cT$ is either balanced on average or fully independent.)
\end{claim}

\begin{proof} 
%
We prove by induction on the depth $h$. When $\cT$ is balanced on average,
we will also show that $\EX_{\cT,x}[f_{h,r}(x)= 1] = 1/2$.
For the base case, set $h=1$. Hence,
all the leaves are directly connected to the root, and the set $\cT$ has only
one function $\tau_r$ (for the root $r$). The probability (over $x$)
that $\tau_r(x) = \tau_r(\flip{x}{\ell})$ is exactly $\Inf_\ell(\tau_r)$.
Suppose $\cT$ is balanced on average.
Since $f_{1,r} = \tau_r$, $\pr_{\cT,x}[f_{1,r}(x)= 1] = \pr_{\cT,x}[\tau_r(x)= 1] = 1/2$.

Now for the induction step. Assume the claim is true for trees of depth $h-1$.
We will denote the indegree of $r$ by $d$.
The root $r$ is connected to a series of subtrees $T_1,T_2,\ldots,T_d$ of depth $h-1$.
The roots of each of these $r_1,r_2,\ldots,r_d$ are the children of $r$.
For convenience, assume that $\ell \in T_1$.
Note that for $b \neq 1$ and $\forall x$, $f_{h-1,r_b}(x) = f_{h-1,r_b}(\flip{x}{\ell})$.
In the final step, the function evaluated is $\tau_r(f_{h-1,r_1}(x), f_{h-1,r_2}(x), \ldots, f_{h-1,r_d}(x))$.

First, let us assume that $\cT$ is fully independent (the proof 
is much easier in this case). The probability that $f_{h-1,r_1}(x) \neq f_{h-1,r_1}(\flip{x}{\ell})$
is, by the induction hypothesis, $\prod_{i = 1}^{h-1} \EX_{\cT}[\Inf_{v_{i-1}}(\tau_{v_{i}})]$.
Conditioned on this, what is the probability that $\tau_v(f_{h-1,r_1}(x), \ldots, )$
$\neq \tau_v(f_{h-1,r_1}(\flip{x}{\ell}),\ldots)$? Since each entry
in the truth table of $\tau_v$ is chosen independently, this
probability is exactly $\EX_{\cT}[\Inf_{v_{h-1}}(\tau_{v_h})]$.
Multiplying, we get that $\EX_\cT[\Inf_\ell(f_{h,r})] = \prod_{i = 1}^{h} \EX_{\cT}[\Inf_{v_{i-1}}(\tau_{v_{i}})]$.
This completes the proof for this case.

Now, we assume that $\cT$ is balanced on average.
For convenience, set random variable $X_i = f_{h-1,r_i}(x)$, and $X'_1$
to be $f_{h-1,r_1}(\flip{x}{\ell})$. For a bit $b$, let
$\cE_i(b)$ denote the event that $X_i = b$. We set $\cF$ to denote
the event that $X_1 \neq X'_1$. We use $\bar{b}$ as shorthand for a vector $b_1,\ldots,b_d$
of bits. The indicator $\chi(\bar{b})$ is $1$
when $\tau_v(\bar{b}) \neq \tau_v(\flip{\bar{b}}{1})$.

\begin{eqnarray*} & & \EX_\cT[\Inf_\ell(f_{h,r})] \\
& = & \EX_\cT[\Pr_x[\tau_v(f_{h-1,r_1}(x), f_{h-1,r_2}(x), \ldots, f_{h-1,r_d}(x)) \\
& & \neq \tau_v(f_{h-1,r_1}(\flip{x}{\ell}), f_{h-1,r_2}(\flip{x}{\ell}), \ldots, f_{h-1,r_d}(x))]] \\
& = & \EX_\cT[\Pr_x[\tau_v(X_1,X_2,\ldots,X_d) \neq \tau_v(X'_1,X_2,\ldots,X_d)]] \\
& = & \EX_\cT[\sum_{\bar{b}} \chi(\bar{b}) \Pr_x[\bigwedge_i \cE_i(b_i) \wedge \cF]] \\
& = & \EX_\cT[\sum_{\bar{b}} \chi(\bar{b}) \Pr_x[\cE_1(b_1) \wedge \cF] \Pr_x[\prod_{i=2}^d \cE_i(b_i)]] \\
& = & \sum_{\bar{b}} \EX_\cT [\chi(\bar{b}) \Pr_x[\cE_1(b_1) \wedge \cF] \Pr_x[\prod_{i=2}^d \cE_i(b_i)]] \\
& = & \sum_{\bar{b}} \EX_\cT [\chi(\bar{b}) \Pr_x[\cE_1(b_1) \wedge \cF]] \Pr_x[\prod_{i=2}^d \EX_\cT[\cE_i(b_i)]] \\
& = & \sum_{\bar{b}} \EX_\cT [\chi(\bar{b}) \Pr_x[\cE_1(b_1) \wedge \cF]] 
\prod_{i=2}^d \EX_\cT[\Pr_x[f_{h-1,r_i} = b_i]] \\
& = & (1/2)^{d-1} \sum_{\bar{b}} \EX_\cT [\chi(\bar{b}) \Pr_x[\cE_1(b_1) \wedge \cF]]
\end{eqnarray*}
The final step uses the induction hypothesis. Now, we use Proposition~\ref{prop:bool}
to deal with $\Pr_x[\cE_1(b_1) \wedge \cF]]$ $= \Pr_x[f_{h-1,r_1}(x) \neq b_1 \wedge (f_{h-1,r_1}(x) = f_{h-1,r_1}(\flip{x}{\ell}))]$.
Let $\cT'$ be the distribution of transfer functions excluding $\tau_r$.
\begin{eqnarray*}
\EX_\cT[\Inf_\ell(f_{h,r})] & = & (1/2)^{d-1} \sum_{\bar{b}} \EX_\cT [\chi(\bar{b}) \Pr_x[\cE_1(b_1) \wedge \cF]]\\
& = & (1/2)^d \sum_{\bar{b}} \EX_\cT[\chi(\bar{b}) \Inf_\ell(f_{h-1,r_1})] \\
& = & \EX_{\cT'} [\Inf_\ell(f_{h-1,r_1})] (1/2)^d \sum_{\bar{b}} \EX_{\cT_r}[\chi(\bar{b})] \\
& = & \prod_{i=1}^{h-1} \EX_{\cT}[\Inf_{v_{i-1}}(\tau_{v_{i}})] \EX_{\cT} [\Pr_{\bar{b}} [\chi(\bar{b})]] \\
& = & \prod_{i=1}^{h-1} \EX_{\cT}[\Inf_{v_{i-1}}(\tau_{v_{i}})] \EX_{\cT} [\Inf_1(\tau_r)] \\
& = & \prod_{i=1}^h \EX_{\cT}[\Inf_{v_{i-1}}(\tau_{v_{i}})]
\end{eqnarray*}
%
%
\qed
\end{proof}

Consider a tree $T$ where all leaves have fixed depth $h$. Set $L$ to be the set
of all leaves of $T$. Define $\Inf(T) = \sum_{\ell \in L} \Inf_\ell(f_{h,r})$.
For a leaf $\ell$ that is a descendant of some vertex $v$, suppose
the path between them is
$v_0 = \ell, v_1, v_2, \ldots, v_a = v$.
Define $\Infprod{\ell}{v}= \prod_{i=1}^{a} \Inf_{v_{i-1}}(\tau_{v_i})$.

\begin{lemma} \label{lem:inf-tree} Let $r_1,r_2,\ldots,r_a$ be the children of the root $r$.
Let $T_i$ be the subtree rooted at $r_i$.
 Then, $\EX_\cT[\Inf(T)] = \sum_i \EX_{\cT_r} [\Inf_{r_i}(\tau_r)] \EX_\cT[\Inf(T_i)]$.
\end{lemma}

\begin{proof} Define $L_i$ to be the set of leaves of $T_i$.
\begin{eqnarray*}
\EX_\cT[\Inf(T)] & = & \EX_\cT[\sum_{\ell \in L} \Inf_\ell(f_{h,r})] \\
& = & \sum_i \EX_\cT[\sum_{\ell \in L_i} \Inf_\ell(f_{h,r})] \\
& = & \sum_i \sum_{\ell \in L_i} \EX_\cT[\Infprod{\ell}{r}] \\
& = & \sum_i \sum_{\ell \in L_i} \EX_{\cT_r}[\Inf_{r_i}(\tau_r)] \EX_\cT[\Infprod{\ell}{r_i}] \\
& = & \sum_i \EX_{\cT_r}[\Inf_{r_i}(\tau_r)] \sum_{\ell \in L_i} \EX_\cT[\Infprod{\ell}{r_i}] \\
& = & \sum_i \EX_{\cT_r}[\Inf_{r_i}(\tau_r)] \EX_\cT[\Inf(T_i)]
\end{eqnarray*}
\qed
\end{proof}

Consider the following randomized tree construction. First, we have a probability
$p_d$, for all positive integers $d$. We define this construction recursively. 
Trees of height $0$ are just singleton vertices. To construct a tree of height $h \geq 1$,
we first have a root $r$. We choose the number of children to be $d$ with
probability $p_d$.
Then, for each child, we recursively construct a tree of height $h-1$.


\begin{lemma} \label{lem:tree} Consider a completely balanced tree $T$ with root $r$ and
 height $h$ generated by the randomized procedure described above.
Furthermore, let the transfer functions be chosen as described above.
Then $\EX[\sum_{\ell \in T} \Inf_\ell(f_{h,r})] = $
$(\sum_{d \geq 1} p_d\I_d)^h$.
\end{lemma}

\begin{proof} For any vertex $v$, let $T_v$ be the subtree rooted at $v$.
 We will show by induction on $h$, that for any vertex at  height $h$,
$\EX[\sum_{\ell \in T_v} \Inf_\ell(f_{h,v})] =$
$(\sum_{d \geq 1} p_d\I_d)^h$.

For vertex $v$ with $h=0$, we trivially have $\Inf_v(f_{0,v}) = 1$.
Now consider $v$ at some height $h$. 
Suppose $v$ has $d$ children $v_1,\ldots,v_d$. For each child $v_i$ of $v$, 
$\EX_{\cT_v}\Inf_{v_i}(\tau_v) = I_d$. 
By the induction hypothesis, 
$\EX[\Inf(T_{v_i})]$ (which are identical for all $i$) has value exactly
$(\sum_{d \geq 1} p_d\I_d)^{h-1}$.
By Lemma~\ref{lem:inf-tree}, conditioned on $v$ having $d$ children
$\EX[\sum_{\ell \in T_v} \Inf_\ell(f_{h,v})]$ 
$= \sum_{i \leq d} \EX_{\cT_r} [\Inf_{v_i}(\tau_v)] \EX[\Inf(T_{v_i})]=$ $I_d(\sum_{d \geq 1} p_d\I_d)^{h-1}$.
By noting that the probability that $v$ has $d$ children is $p_d$, $\EX[\sum_{\ell \in T_v} \Inf_\ell(f_{h,v})] =$
$(\sum_{d \geq 1} p_d\I_d)^h$.
%
\qed
\end{proof}

\section{The topology of the Kauffman network} \label{sec:top}

We now prove some topological properties of the random graphs (which
are effectively directed Erd\H{o}s-R\'{e}nyi graphs). We remind that reader
that the indegrees for all vertices are chosen independently from the same
distribution.

\begin{definition} \label{def:nbd} The \emph{distance $t$ in-neighborhood of $i$} 
is denoted by $N^-_{t,i}$. This is the set of all vertices whose shortest
path distance (along directed paths) to $i$ is exactly $t$. We set
$\nbdt{t}{i} = \bigcup_{s \leq t} \nbd{s}{i}$.

We define $t^* = \log n/(4\log K_{max})$.
\end{definition}

\begin{claim} \label{clm:g-tree} Fix a vertex $i$ and let
$t \leq t^*$. Let $\cE_i$ denote the event that 
the subgraph induced by $\nbdt{t^*}{i}$ is a directed tree
with edges directed towards root $i$. Then, 
$|\nbdt{t^*}{i}| \leq n^{1/4}$ and
$\pr(\cE_i) \geq 1 - 1/\sqrt{n}$.
\end{claim}

\begin{proof} Let us start with an empty graph, and slowly add random edges
in a prescribed order. We begin with $i$, and then choose the incoming edges.
This gives the set $\nbd{1}{i}$. Then, we choose all the in-neighbors of $\nbd{1}{i}$.
This is done by iterating over all vertices in $\nbd{1}{i}$, and for each such vertex,
selecting every vertex as a neighbor with probability $K/n$.
This gives us $\nbd{2}{i}$. Proceeding this way, we incrementally build up $\nbdt{t^*}{i}$.
Note that $|\nbdt{t^*}{i}| \leq K_{max}^{t^*} = n^{1/4}$.

Consider the construction of $N^-_{\leq t,i}$. Every new element added
to this set is a uniform random element from $[n]$. 
Consider a random sequence of $n^{1/4}$ elements chosen
uniformly at random (with replacement) from $[n]$.
The probability that no element is repeated at least 
$$ \left(1 - \frac{n^{1/4}}{n}\right)^{n^{1/4}} \geq \exp(-\sqrt{n}/n) \geq 1 - (\sqrt{n})^{-1} $$
If no element in $N^-_{\leq t,i}$ is repeated, then the subgraph 
induced by $N^-_{\leq t,i}$ is a directed tree.
\qed
\end{proof}

\section{Proof of the Main Theorem} \label{sec:final}

We are now ready to prove our main result.

\begin{proof} (of Theorem~\ref{thm:avg-inf}) By Claim~\ref{clm:avg-inf}, the average influence
at time $t$ is $(\sum_i \sum_j\Inf_j(f_{t,i}))/n$.
For a fixed vertex $i$, let us compute $\sum_j \Inf_j(f_{t,i})$.
Denote this quantity by $X$. We apply Bayes rule 
to split $\EX[X]$ into conditional expectations.
$$ \EX[X] = \pr(\cE_i) \EX[X | \cE_i] + \pr(\overline{\cE_i}) \EX[X | \overline{\cE_i}]$$
Observing that $X$ is always positive and applying 
Claim~\ref{clm:g-tree}, we get $\EX[X] \geq (1-1/\sqrt{n}) \EX[X | \cE_i]$.
Note that since $ \Inf_j(f_{t,i}) \leq 1$, $X \leq |\nbd{t}{i}| \leq n^{1/4}$.
Hence $\EX[X] \geq \EX[X | \cE_i] - n^{-1/4}$. We now obtain
an upper bound applying Claim~\ref{clm:g-tree} again.
\begin{eqnarray*}
\EX[X] & \leq & \pr(\cE_i) \EX[X | \cE_i] 
+ (1/\sqrt{n}) \EX[X | \overline{\cE_i}] \\
& \leq & \EX[X | \cE_i] + n^{-1/4}
\end{eqnarray*}
%
It only remains to determine $\EX[X | \cE_i]$. Conditioned on $\cE_i$, the
induced subgraph on $\nbdt{t}{i}$ is a directed tree.
We apply Lemma~\ref{lem:tree}
to complete the proof.
\qed\\
\end{proof}

\end{document}